\documentstyle[12pt,epsfig]{article}
\date{}
\begin{document}

{\large \sf
\title{\vspace{2cm}
{\LARGE \sf
Hidden Symmetry of the CKM and \\
Neutrino Mapping Matrices} \vspace{3cm}}

{\large \sf
\author{
{\large \sf
R. Friedberg$^1$ and  T. D. Lee$^{1,~2}$}\\
{\normalsize \it 1. Physics Department, Columbia University}\\
{\normalsize \it New York, NY 10027, U.S.A.}\\
{\normalsize \it 2. China Center of Advanced Science and Technology (CCAST/World Lab.)}\\
{\normalsize \it P.O. Box 8730, Beijing 100080, China}\\
} \maketitle

\newpage

\begin{abstract}

{\normalsize \sf

We propose that the smallness of the light quark masses is related
to the smallness of the $T$ (i.e. $CP$) violation in hadronic weak
interactions. Accordingly, for each of the two quark sectors
("upper" and "lower") we construct a $3\times 3$ mass matrix in a
bases of unobserved quark states, such that the "upper"and "lower"
basis states correspond exactly via the $W^\pm$ transitions in the
weak interaction. In the zeroth approximation of our formulation,
we assume $T$ conservation by making all matrix elements real. In
addition, we impose a "hidden symmetry" (invariance under
simultaneous translations of all three basis quark states in each
sector), which ensures a zero mass eigenstate in each sector.

Next, we simultaneously break the hidden symmetry and $T$
invariance by introducing a phase factor $e^{i\chi}$ in the
interaction for each sector. The Jarlskog invariant $J_{CKM}$, as
well as the light quark masses are evaluated in terms of the
parameters of the model. Comparing formulas, we find that most
unknown factors drop out, resulting in a simple relation with
$J_{CKM}=(m_dm_s/m_b^2)^{1/2}A\lambda^3\cos\frac{1}{2}\chi$, to
leading order in $\chi$ and $m_s/m_b$, with $A,~\lambda$ the
Wolfenstein parameters. (Because of the large top quark mass, the
contribution from upper quark sector can be neglected.) Setting
$J_{CKM}=3.08\times 10^{-5}$, $m_b=4.7GeV$ ($1s$ mass),
$m_s=95MeV$, $A=0.818$ and $\lambda=0.227$, we find
$m_d\cos^2\frac{1}{2}\chi \simeq 2.4MeV$, consistent with the
accepted value $m_d=3-7MeV$.

We make a parallel proposal for the lepton sectors. With the
hidden symmetry and in the approximation of $T$ invariance, both
the masses of $e$ and $\nu_1$ are zero. The neutrino mapping
matrix $V_\nu$ is shown to be of the same Harrison-Scott form
which is in agreement with experiments. We also examine the
correction due to $T$ violation, and evaluate the corresponding
Jarlskog invariant ${\cal J}_\nu$.

 }
\end{abstract}

\vspace{1cm}

{\normalsize \sf PACS{:~~14.60.Pq,~~11.30.Er}}

\vspace{1cm}


\newpage

\section*{\Large \sf  1. Introduction}
\setcounter{section}{1} \setcounter{equation}{0}

In a recent paper[1], we postulate a new symmetry of the neutrino
mass matrix in terms of the field operators $\nu_e,~\nu_\mu$ and
$\nu_\tau$. This symmetry enables us to derive the Harrison-Scott
form[2,3] of the neutrino mapping matrix $V_\nu$. However, the
formalism has a built-in asymmetry between the charged leptons and
the neutral ones. In this paper, we modify the symmetry introduced
in [1], as that $e$, $\mu$, $\tau$ and $\nu_1,~\nu_2$, $\nu_3$ are
now set on a similar basis. Furthermore, the new symmetry can also
be extended to quarks $d$, $s$, $b$ and $u$, $c$, $t$. For
clarity, we first discuss how the new symmetry, called hidden
symmetry, can be realized in the quark sectors leading to the CKM
matrix $U_{CKM}$. Next, we discuss its application to leptons,
resulting again in the Harrison-Scott form of the neutrino mapping
matrix $V_\nu$.

In the quark sector, let $q_i(\downarrow)$ and $q_i(\uparrow)$ be
the quark states "diagonal" in $W^\pm$ transitions:
$$
q_i(\downarrow)\rightleftharpoons q_i(\uparrow) + W^-
$$
$${\sf and}~~~~~~~~~~~~~~~~~~~~~~~~~~~~~~~~~~~~~~~~~
~~~~~~~~~~~~~~~~~~~~~~~~~~~~~~~~~~~~~~~\eqno(1.1)$$
$$
q_i(\uparrow)\rightleftharpoons q_i(\downarrow) + W^+
$$
with $i=1,~2,~3$. Their electric charges in units of $e$ are
$-\frac{1}{3}$ for $q_i(\downarrow)$, and $+\frac{2}{3}$ for
$q_i(\uparrow)$. However, these are not the observed mass
eigenstates $d$, $s$, $b$ and $u$, $c$, $t$. Likewise, let
$l_i(\downarrow)$ and $l_i(\uparrow)$ be the lepton states
"diagonal" in the corresponding $W^\pm$ transitions:
$$
l_i(\downarrow)\rightleftharpoons l_i(\uparrow) + W^-
$$
$${\sf and}~~~~~~~~~~~~~~~~~~~~~~~~~~~~~~~~~~~~~~~~~~~~~~~~~~~~~~\eqno(1.2)$$
$$
l_i(\uparrow)\rightleftharpoons l_i(\downarrow) + W^+
$$
with their electric charge unit $-1$ for $l_i(\downarrow)$, and
$0$ for $l_i(\uparrow)$ and $i=1,~2,~3$. Again, neither
$l_i(\downarrow)$ nor $l_i(\uparrow)$ are the mass eigenstates
$e$, $\mu$, $\tau$ and $\nu_1,~\nu_2$, $\nu_3$. Thus, for each of
these four triplets
$$
\{q_1(\downarrow),~q_2(\downarrow),~q_3(\downarrow)\},~~\{
q_1(\uparrow),~q_2(\uparrow),~q_3(\uparrow)\}
$$
$$
\{l_1(\downarrow),~l_2(\downarrow),~l_3(\downarrow)\}~~{\sf
and}~~\{ l_1(\uparrow),~l_2(\uparrow),~l_3(\uparrow)\}\eqno(1.3)
$$
there exists a separate $3\times 3$ mass matrix, denoted by
$$
M(q_\downarrow),~M(q_\uparrow),~M(l_\downarrow)~~{\sf and}~~
M(l_\uparrow)\eqno(1.4)
$$
respectively. As we shall discuss, these mass matrices satisfy a
common set of rules due to hidden symmetry, leading to a unifying
formalism of both the CKM matrix $U_{CKM}$ and the neutrino
mapping matrix $V_\nu$.

In what follows, we begin our discussion in the approximation
assuming time reversal invariance $T$. Thus, the mass matrix
$M(q_\downarrow)$, $M(q_\uparrow)$, $M(l_\downarrow)$ and
$M(l_\uparrow)$ are all $3\times 3$ real symmetric matrices. The
corresponding mass operators are
$$
{\cal
M}(q_\downarrow)=\bigg(\bar{q}_1(\downarrow),~\bar{q}_2(\downarrow),~\bar{q}_3(\downarrow)\bigg)
M(q_\downarrow) \left(
\begin{array}{r}
q_1(\downarrow)\\
q_2(\downarrow)\\
q_3(\downarrow)
\end{array}\right), \eqno(1.5)
$$
$$
{\cal
M}(q_\uparrow)=\bigg(\bar{q}_1(\uparrow),~\bar{q}_2(\uparrow),~\bar{q}_3(\uparrow)\bigg)
M(q_\uparrow) \left(
\begin{array}{r}
q_1(\uparrow)\\
q_2(\uparrow)\\
q_3(\uparrow)
\end{array}\right), \eqno(1.6)
$$
$$
{\cal
M}(l_\downarrow)=\bigg(\bar{l}_1(\downarrow),~\bar{l}_2(\downarrow),~\bar{l}_3(\downarrow)\bigg)
M(l_\downarrow) \left(
\begin{array}{r}
l_1(\downarrow)\\
l_2(\downarrow)\\
l_3(\downarrow)
\end{array}\right), \eqno(1.7)
$$
and
$$
{\cal
M}(l_\uparrow)=\bigg(\bar{l}_1(\uparrow),~\bar{l}_2(\uparrow),~\bar{l}_3(\uparrow)\bigg)
M(l_\uparrow) \left(
\begin{array}{r}
l_1(\uparrow)\\
l_2(\uparrow)\\
l_3(\uparrow)
\end{array}\right).\eqno(1.8)
$$
In (1.5), $q_i(\downarrow)$ and $\bar{q}_i(\downarrow)$ are
related to the corresponding Dirac field operators
$\psi(q_i(\downarrow))$ and its Hermitian conjugate
$\psi^\dag(q_i(\downarrow))$ by
$$
q_i(\downarrow)=\psi(q_i(\downarrow))~~{\sf
and}~~\bar{q}_i(\downarrow)=\psi^\dag(q_i(\downarrow))\gamma_4.\eqno(1.9)
$$
Likewise, in(1.6)-(1.8) $q_i(\uparrow)$, $\bar{q}_i(\uparrow)$,
$l_i(\downarrow)$, etc. are similarly related to their
corresponding Dirac field operators. We assume that each of these
four mass operators (1.5)-(1.8) satisfies a hidden symmetry with
${\cal M}(q_\downarrow)$ invariant under the transformation
$$
q_1(\downarrow)\rightarrow q_1(\downarrow)+z,
~~q_2(\downarrow)\rightarrow q_2(\downarrow)+\eta_\downarrow z
~~{\sf  and}~~ q_3(\downarrow)\rightarrow
q_3(\downarrow)+\xi_\downarrow\eta_\downarrow z \eqno(1.10)
$$
where $z$ is a space-time independent constant element of the
Grassmann algebra anticommuting with the Dirac field operators,
and $\xi_\downarrow$, $\eta_\downarrow$ are $c$-numbers. It will
be shown in the next section that (1.10) implies a zero down-quark
mass in the absence of $T$ violation. Similar symmetries are also
assumed for other triplets $\{q_i(\uparrow)\}$,
$\{l_i(\downarrow)\}$ and $\{l_i(\uparrow)\}$. Thus, we correlate
the nearly zero masses of $d$, $u$. $e$ and $\nu_1$ with $T$
invariance and the new symmetry.

In Section~3. we derive the form of CKM matrix in the same zeroth
approximation of $T$ invariance. The violation of $T$ invariance
will be discussed in Section~4. As will be shown, to the first
approximation of small $T$ violation, we derive an interesting
formula relating $T$ violating Jarlskog invariant $J$ with quark
masses:
$$
J = \bigg(\frac{m_d m_s}{m_b^2} \bigg)^{\frac{1}{2}} A \lambda^3
\cos [ {\scriptstyle {1\over 2}}{\displaystyle
\chi}_{\scriptscriptstyle T}(\downarrow)] + O \bigg(\frac{m_u
m_c}{m_t^2}\bigg)^{\frac{1}{2}} \eqno(1.11)
$$
where[4] $A\cong0.818$, $\lambda\cong 0.2272$ are the Wolfenstein
parameters. Using the experimental values[4] $J\cong 3.08 \cdot
10^{-5}$, $m_s\cong 95MeV$ and $m_b\cong 4.7GeV$ ($1s$ mass), we
find
$$
m_d \cos^2[{\scriptstyle \frac{1}{2}}{\displaystyle
\chi}_{\scriptscriptstyle T}(\downarrow)]\cong 2.4MeV,\eqno(1.12)
$$
where ${\displaystyle \chi}_{\scriptscriptstyle T}(\downarrow)$ is
the $T$-violating phase in the $\downarrow$ quark sector. Since
$\cos\frac{1}{2}{\displaystyle \chi}_{\scriptscriptstyle
T}(\downarrow)$$\leq$$1$, we have
$$
m_d\geq 2.4 MeV\eqno(1.13)
$$
consistent with the range $m_d\cong 3$ to $7MeV$ quoted by the
particle data group[4].

In section~5 and 6, we discuss lepton sectors. As in Ref.[1], we
show how the new hidden symmetry can also lead to the
Harrison-Scott form of the neutrino-mapping matrix $V_\nu$ in
agreement with experiments. The Jarlskog invariant in the lepton
sector is also calculated to the lowest order of the $T$ violating
interaction.

\newpage

\section*{\Large \sf  2. Hidden Symmetry}
\setcounter{section}{2} \setcounter{equation}{0}

Consider first the $\{q_i(\downarrow)\}$ sector. In the
approximation of $T$ invariance, the $3\times 3$ matrix in (1.5)
becomes a real symmetric matrix $M_0(q_\downarrow)$ characterized
by six real parameters: three diagonal and three off-diagonal
elements. We propose to represent the corresponding mass operator
${\cal M}(q_\downarrow)$ by
$$
{\cal M}_0(q_\downarrow) = \alpha_\downarrow | q_3(\downarrow) -
\xi_\downarrow q_2(\downarrow)|^2 + \beta_\downarrow |
q_2(\downarrow) - \eta_\downarrow q_1(\downarrow)|^2
+\gamma_\downarrow |q_1(\downarrow) - \zeta_\downarrow
q_3(\downarrow)|^2\eqno(2.1)
$$
with also six real parameters
$\alpha_\downarrow,~\beta_\downarrow,~\gamma_\downarrow,
~\xi_\downarrow,~\eta_\downarrow$ and $\zeta_\downarrow$. Their
relation with the six diagonal and off-diagonal elements of an
arbitrary symmetric matrix $M(q)$ is given in the Appendix. We
impose the hidden symmetry requirement that ${\cal
M}_0(q_\downarrow)$ be invariant under the transformation (1.10).
Substituting (1.10) into (2.1) and requiring the symmetry, we see
that these three parameters $\xi_\downarrow,~\eta_\downarrow$ and
$\zeta_\downarrow$ must satisfy
$$
\xi_\downarrow\eta_\downarrow\zeta_\downarrow=1.\eqno(2.2)
$$
The corresponding mass matrix $M(q_\downarrow)$ defined by (1.5)
is
$$
M_0(q_\downarrow) = \left(
\begin{array}{ccc}
\gamma+\beta\eta^2 & -\beta\eta & -\gamma\zeta\\
-\beta\eta &\beta + \alpha\xi^2 & -\alpha\xi\\
-\gamma\zeta& -\alpha\xi & \alpha+\gamma\zeta^2
\end{array}
\right)_\downarrow \eqno(2.3)
$$
where the suffix $\downarrow$ on the right hand side indicates
that the parameters $\alpha,~\beta,~\gamma$, $\xi,~\eta,~\zeta$
refer to $\alpha_\downarrow,~\beta_\downarrow,
~\gamma_\downarrow,~\xi_\downarrow,~\eta_\downarrow$ and
$\zeta_\downarrow$ respectively. From (2.3), we see that the
determinant of $M_0(q_\downarrow)$ is given by
$$
|M_0(q_\downarrow)|=\bigg[\alpha\beta\gamma(\xi\eta\zeta-1)^2\bigg]_\downarrow~.\eqno(2.4)
$$
Thus, for
$$
(\xi\eta\zeta)_\downarrow\equiv
\xi_\downarrow\eta_\downarrow\zeta_\downarrow=1.\eqno(2.5)
$$
we have
$$
|M_0(q_\downarrow)|=0.\eqno(2.6)
$$
Choose $\alpha_\downarrow,~\beta_\downarrow$ and
$\gamma_\downarrow$ to be all positive. The operator ${\cal
M}_0(q_i)$ is then positive; condition (2.5) implies the smallest
eigenvalue of $M(q_\downarrow)$, the down quark mass, to be zero;
i.e.,
$$
m_d=0\eqno(2.7)
$$
on account of the hidden symmetry requirement (1.10) and the
approximation of $T$ invariance.

This result can also be seen directly from the symmetry
requirement (1.10). In the three-dimensional space of coordinate
axes $q_1(\downarrow),~q_2(\downarrow)$ and $q_3(\downarrow)$, the
transformation (1.10) represents a translation along the direction
parallel to the three dimensional unit vector
$$
\epsilon_\downarrow \propto \left(
\begin{array}{l}
1\\
\eta_\downarrow\\
\xi_\downarrow \eta_\downarrow
\end{array}
\right). \eqno(2.8)
$$
The assumed invariance under (1.10) is identical to the invariance
of ${\cal M}_0(q_\downarrow)$ under a translation along the vector
$\epsilon_\downarrow$; thus, $\epsilon_\downarrow$ is an
eigenvector of the corresponding mass matrix $M_0(q_\downarrow)$,
with zero eigenvalue (i.e., zero mass).

Likewise, under the transformation
$\downarrow\rightarrow\uparrow$,
$$
(\alpha_\downarrow,~\beta_\downarrow,~\gamma_\downarrow)\rightarrow
(\alpha_\uparrow,~\beta_\uparrow,~\gamma_\uparrow)
$$
$${\sf and}~~~~~~~~~~~~~~~~~~~~~~~~~~~~~~~~~~~~~~~~~~~~~~~~~~~~~~\eqno(2.9)$$
$$
(\xi_\downarrow,~\eta_\downarrow,~\zeta_\downarrow)\rightarrow
(\xi_\uparrow,~\eta_\uparrow,~\zeta_\uparrow)
$$
we have
$$
M_0(q_\downarrow)\rightarrow M_0(q_\uparrow)~~{\sf and}~~{\cal
M}_0(q_\downarrow)\rightarrow {\cal M}_0(q_\uparrow). \eqno(2.10)
$$
As in (1.10), the hidden symmetry
$$
q_1(\uparrow)\rightarrow
q_1(\uparrow)+z,~~q_2(\uparrow)\rightarrow
q_2(\uparrow)+\eta_\uparrow z~~{\sf and}~~q_3(\uparrow)\rightarrow
q_3(\uparrow)+\xi_\uparrow\eta_\uparrow z \eqno(2.11)
$$
implies the corresponding invariance of the mass operator ${\cal
M}_0(q_\uparrow)$, in the approximation of $T$ invariance. Thus,
(2.11) implies
$$
\xi_\uparrow\eta_\uparrow\zeta_\uparrow=1\eqno(2.12)
$$
and the up quark $u$ to be of zero mass; i.e., with hidden
symmetry and $T$ invariance,
$$
m_u=0.\eqno(2.13)
$$

\newpage

\section*{\Large \sf  3. CKM Matrix (neglecting $T$ violation)}
\setcounter{section}{3} \setcounter{equation}{0}

In  this section, we discuss the CKM matrix in the same zeroth
approximation by neglecting $T$ violation. Let $(U_\downarrow)_0$
and $(U_\uparrow)_0$ be the unitary matrices that diagonalize
$M_0(q_\downarrow)$ and $M_0(q_\uparrow)$:
$$
(U_\downarrow)_0^\dag M_0(q_\downarrow)(U_\downarrow)_0 =\left(
\begin{array}{ccc}
m_0(d)&0&0\\
0&m_0(s)&0\\
0&0&m_0(b)
\end{array}
\right )\eqno(3.1)
$$
and
$$
(U_\uparrow)_0^\dag M_0(q_\uparrow)(U_\uparrow)_0 =\left(
\begin{array}{ccc}
m_0(u)&0&0\\
0&m_0(c)&0\\
0&0&m_0(t)
\end{array}
\right ).\eqno(3.2)
$$
The corresponding CKM matrix is given by
$$
(U_{CKM})_0=(U_\uparrow)_0^\dag(U_\downarrow)_0.\eqno(3.3)
$$
In accordance with (2.7) and (2.13), we have in the notation of
(3.1) and (3.2)
$$
m_0(d)=m_0(u)=0.\eqno(3.4)
$$
Without $T$ violation, $(U_\downarrow)_0$, $(U_\uparrow)_0$ and
$(U_{CKM})_0$ are each a $3\times 3$ real orthogonal matrix
characterized by three real parameters.

In (2.3), the mass matrix $M_0(q_\downarrow)$ has six parameters
$\alpha_\downarrow,~\beta_\downarrow,
~\gamma_\downarrow,~\xi_\downarrow,~\eta_\downarrow$ and
$\zeta_\downarrow$. With the constraint
$\xi_\downarrow\eta_\downarrow\zeta_\downarrow=1$ in accordance
with (2.5), there are still five independent parameters in
$M_0(q_\downarrow)$. Together with $M_0(q_\uparrow)$, we have
$5+5=10$ parameters. Assuming that the only observables are the
quark masses and the CKM matrix. Since $m_0(d)=m_0(u)=0$ in this
approximation, there are only four nonzero masses
$m_0(s),~m_0(b),~m_0(c)$ and $m_0(t)$. In addition, the CKM matrix
with $T$ invariance is characterized by three real parameters;
together, there are
$$
4+3=7
$$
observables in this approximation. That means among the $10$
parameters, there are
$$
10-7=3\eqno(3.5)
$$
parameters which are "unphysical". The elimination of these three
unphysical parameters is analogous to the gauge condition in a
vector field theory. As we shall see, a convenient choice is to
eliminate two of these three by requiring
$$
\frac{\beta_\downarrow}{\gamma_\downarrow}=\zeta_\downarrow^2~~{\sf
and}~~\frac{\beta_\uparrow}{\gamma_\uparrow}=\zeta_\uparrow^2.\eqno(3.6)
$$

Define four real angular variables
$\theta_\downarrow,~\phi_\downarrow$ and
$\theta_\uparrow,~\phi_\uparrow$ by
$$
\xi_\downarrow=\tan \phi_\downarrow,~~\xi_\uparrow=\tan
\phi_\uparrow
$$
$$
\eta_\downarrow=\tan \theta_\downarrow \cos \phi_\downarrow~~{\sf
and}~~\eta_\uparrow=\tan \theta_\uparrow \cos \phi_\uparrow.
\eqno(3.7)
$$
It can be readily verified that with (3.6) the eigenstates of
$M_0(q_\downarrow)$ become quite simple, given by
$$
\epsilon_\downarrow = \left(
\begin{array}{l}
\cos \theta_\downarrow\\
\sin \theta_\downarrow \cos \phi_\downarrow\\
\sin \theta_\downarrow \sin \phi_\downarrow
\end{array}
\right)~{\sf with~eigenvalue}~\lambda(\epsilon_\downarrow),
\eqno(3.8)
$$
$$
p_\downarrow = \left(
\begin{array}{l}
-\sin \theta_\downarrow\\
\cos \theta_\downarrow \cos \phi_\downarrow\\
\cos \theta_\downarrow \sin \phi_\downarrow
\end{array}
\right)~{\sf with~eigenvalue}~\lambda(p_\downarrow), \eqno(3.9)
$$
and
$$
P_\downarrow = \left(
\begin{array}{l}
0\\
-\sin \phi_\downarrow\\
\cos \phi_\downarrow
\end{array}
\right)~{\sf with~eigenvalue}~\lambda(P_\downarrow). \eqno(3.10)
$$
Here $\lambda(\epsilon_\downarrow),~\lambda(p_\downarrow)$ and
$\lambda(P_\downarrow)$ are the same $0$th order approximation
$m_0(d),~m_0(s)$ and $m_0(b)$ in (3.1), with
$$
m_0(d)=\lambda(\epsilon_\downarrow)=0,\eqno(3.11)
$$
$$
m_0(s)=\lambda(p_\downarrow)=\beta_\downarrow[1+\eta_\downarrow^2(1+\xi_\downarrow^2)]\eqno(3.12)
$$
and
$$
m_0(b)=\lambda(P_\downarrow)=\alpha_\downarrow(1+\xi_\downarrow^2)+\beta_\downarrow.\eqno(3.13)
$$
In terms of $\xi_\downarrow$ and $\eta_\downarrow$, the
statevector $\epsilon_\downarrow$ satisfies (2.8).

Likewise, the eigenstates of $M_0(q_\uparrow)$ are
$$
\epsilon_\uparrow = \left(
\begin{array}{l}
\cos \theta_\uparrow\\
\sin \theta_\uparrow \cos \phi_\uparrow\\
\sin \theta_\uparrow \sin \phi_\uparrow
\end{array}
\right)~{\sf with~eigenvalue}~\lambda(\epsilon_\uparrow),
\eqno(3.14)
$$
$$
p_\uparrow = \left(
\begin{array}{l}
-\sin \theta_\uparrow\\
\cos \theta_\uparrow \cos \phi_\uparrow\\
\cos \theta_\uparrow \sin \phi_\uparrow
\end{array}
\right)~{\sf with~eigenvalue}~\lambda(p_\uparrow), \eqno(3.15)
$$
and
$$
P_\uparrow = \left(
\begin{array}{l}
0\\
-\sin \phi_\uparrow\\
\cos \phi_\uparrow
\end{array}
\right)~{\sf with~eigenvalue}~\lambda(P_\uparrow) \eqno(3.16)
$$
where
$$
m_0(u)=\lambda(\epsilon_\uparrow)=0,\eqno(3.17)
$$
$$
m_0(c)=\lambda(p_\uparrow)=\beta_\uparrow[1+\eta_\uparrow^2(1+\xi_\uparrow^2)]\eqno(3.18)
$$
and
$$
m_0(t)=\lambda(P_\uparrow)=\alpha_\uparrow(1+\xi_\uparrow^2)+\beta_\uparrow.\eqno(3.19)
$$
Correspondingly, the $3\times 3$ unitary matrices
$(U_\downarrow)_0$ and $(U_\uparrow)_0$ of (3.1) and (3.2) are
given by
$$
(U_\downarrow)_0=(\epsilon_\downarrow,~p_\downarrow,~P_\downarrow)\eqno(3.20)
$$
and
$$
(U_\uparrow)_0=(\epsilon_\uparrow,~p_\uparrow,~P_\uparrow).\eqno(3.21)
$$
Thus, in accordance with (3.3), the corresponding CKM matrix in
the same approximation is given by
$$
(U_{CKM})_0=\left(
\begin{array}{ccc}
\cos \theta_\downarrow\cos \theta_\uparrow &-\sin\theta_\downarrow \cos\theta_\uparrow &\sin\theta_\uparrow \sin\phi\\
~~+\sin\theta_\downarrow \sin\theta_\uparrow \cos\phi
&~~+\cos\theta_\downarrow \sin\theta_\uparrow \cos\phi &\\
&&\\
-\cos \theta_\downarrow\sin \theta_\uparrow &\sin\theta_\downarrow \sin\theta_\uparrow &\cos\theta_\uparrow \sin\phi\\
~~+\sin\theta_\downarrow \cos\theta_\uparrow \cos\phi
&~~+\cos\theta_\downarrow \cos\theta_\uparrow \cos\phi &\\
&&\\
 -\sin\theta_\downarrow \sin\phi
&-\cos\theta_\downarrow\sin\phi&\cos\phi
\end{array}
 \right )~,\eqno(3.22)
$$
in which
$$
\phi=\phi_\uparrow-\phi_\downarrow.\eqno(3.23)
$$
Equations (3.6) and (3.23) eliminate the three unphysical
variables, as we shall see. Upon comparison with experimental
values, we find from (3.22)
$$
\theta_\uparrow-\theta_\downarrow={\sf Cabibbo~angle}\eqno(3.24)
$$
with
$$
\sin(\theta_\uparrow-\theta_\downarrow)\cong\lambda=0.227.\eqno(3.25)
$$
By taking the ratio of (1,3) and (2,3) matrix elements of
$(U_{CKM})_0$, we estimate $\theta_\uparrow=O(\lambda)$; likewise,
from the corresponding (3,1) and (3,2) matrix elements,
$\theta_\downarrow=O(\lambda)$. Using the (2,3) matrix element, we
derive
$$
\sin\phi\cong A\lambda^2\eqno(3.26)
$$
with $A=0.818$.

We observe that the dependence of $(U_{CKM})_0$ on
$\phi_\downarrow$ and $\phi_\uparrow$ is only through
$\phi=\phi_\uparrow-\phi_\downarrow$. Thus, $(U_{CKM})_0$ is
independent of $\phi_\uparrow+\phi_\downarrow$, which together
with the two conditions given by (3.6) eliminate the 3 unphysical
parameters mentioned in (3.5).

\newpage

\section*{\Large \sf  4. $T$-Violation}
\setcounter{section}{4} \setcounter{equation}{0}

In the approximation of $T$ invariance, by using (2.1) and
constraints
$$
\xi_\downarrow \eta_\downarrow \zeta_\downarrow = 1~~~~{\rm
and}~~~ {\beta_\downarrow \over \gamma_\downarrow} =
\zeta_\downarrow^2\eqno(4.1)
$$
in accordance with (2.5) and (3.6), we find that the mass operator
(2.1) can also be written as
$$
{\cal M}_0(q_\downarrow) = \alpha_\downarrow | q_3(\downarrow) -
\xi_\downarrow q_2(\downarrow)|^2 + \beta_\downarrow |
q_2(\downarrow) - \eta_\downarrow q_1(\downarrow)|^2
+\beta_\downarrow |q_3(\downarrow) - \xi_\downarrow\eta_\downarrow
q_1(\downarrow)|^2\eqno(4.2)
$$
With $T$ violation, we replace ${\cal M}_0(q_\downarrow)$ by
$$
{\cal M}(q_\downarrow) = \alpha_\downarrow | q_3(\downarrow) -
\xi_\downarrow e^{i{\displaystyle \chi}_{\scriptscriptstyle T}}
q_2(\downarrow)|^2 + \beta_\downarrow | q_2(\downarrow) -
\eta_\downarrow q_1(\downarrow)|^2 +\beta_\downarrow
|q_3(\downarrow) - \xi_\downarrow\eta_\downarrow
q_1(\downarrow)|^2\eqno(4.3)
$$
in which
$$
{\displaystyle \chi}_{\scriptscriptstyle T} = {\displaystyle
\chi}_{\scriptscriptstyle T} (\downarrow)\eqno(4.4)
$$
is the $T$-violating phase factor for  the $\downarrow$ quark
sector. The corresponding mass matrix defined by (1.5) is given by
$$
 M(q_\downarrow)=  M_0(q_\downarrow)
+ M_1(q_\downarrow)\eqno(4.5)
$$
with $M_0(q_\downarrow)$ given by (2.3). Because of (2.5) and the
first equation in (3.6), $M_0(q_\downarrow)$ can also be written
as
$$
M_0(q_\downarrow) = \left(
\begin{array}{ccc}
\beta\eta^2(1+\xi^2) & -\beta\eta & -\beta\xi\eta\\
-\beta\eta &\beta + \alpha\xi^2 & -\alpha\xi\\
-\beta\xi\eta & -\alpha\xi & \alpha+\beta
\end{array}
\right)_\downarrow~. \eqno(4.6)
$$
The $T$ violating term in (4.5) is
$$
M_1(q_\downarrow) = \alpha_\downarrow \xi_\downarrow \left(
\begin{array}{ccc}
~~0 ~~& 0 & 0\\
0 & 0 & 1-e^{-i{\displaystyle \chi}_{\scriptscriptstyle T}(\downarrow)}\\
0 &  1-e^{i{\displaystyle \chi}_{\scriptscriptstyle T}(\downarrow)}& 0\\
\end{array}
\right)~. \eqno(4.7)
$$
Because of $T$ violation, the mass of $d$ quark is not zero and
the CKM matrix is unitary but not real.

\noindent {\bf 4.1 $d$ quark mass}

In accordance with (3.11)-(3.13), the eigenvalues of
$M_0(q_\downarrow)$ are $\lambda (\epsilon_\downarrow) = 0$,
$\lambda(p_\downarrow)$ and $\lambda(P_\downarrow)$, whereas those
of $M (q_\downarrow)$ are the observed quark masses $m=m_d,~m_s~$
and $m_b$ determined by
$$
|M (q_\downarrow) - m | = 0. \eqno(4.8)
$$
By using (3.11)-(3.13) and (4.5)-(4.7), we find (4.8) to be the
cubic equation,
$$
m(m-\lambda(p_\downarrow))(m - \lambda(P_\downarrow))= | M
(q_\downarrow)|\eqno(4.9)
$$
with
$$
|M (q_\downarrow)| =2
\alpha_\downarrow\beta_\downarrow^2\xi_\downarrow^2\eta_\downarrow^2
(1-\cos {\displaystyle \chi}_{\scriptscriptstyle T}(\downarrow)).
\eqno(4.10)
$$
Since in accordance with (3.12)-(3.13), $\lambda_(p_\downarrow) =
m_0(s)$ and $\lambda(P_\downarrow) = m_0(b)$ are the zeroth order
values of $m_s$ and $m_b$, both are $\gg m_d$. From (3.7) and
(3.12)-(3.13),
$$
\begin{array}{lll}
&m_0(s) &= \lambda (p_\downarrow) =\beta_\downarrow \sec^2 \theta_\downarrow\\
{\sf and}~~~~~~~~~&&\\
&m_0(b) &= \lambda (P_\downarrow) =\alpha_\downarrow \sec^2 \phi_\downarrow+\beta_\downarrow.\\
\end{array}\eqno(4.11)
$$
By setting $m=m_d,~m_s$ and $m_b$ respectively in (4.9), we have
$$
m_d = \left[ (m_0(s) - m_d)(m_0(b) - m_d)\right]^{-1}
|M(q_\downarrow)|, \eqno(4.12)
$$
$$
m_s-m_0(s)  = - \left[ m_s (m_0(b) - m_s)\right]^{-1}
|M(q_\downarrow)|\eqno(4.13)
$$
and
$$
m_b-m_0(b)  = \left[ m_b (m_b - m_0(s))\right]^{-1}
|M(q_\downarrow)|.\eqno(4.14)
$$
Thus, neglecting corrections $O(m_d/m_s)$ and $O(m_d/m_b)$, we
find from (4.12)-(4.14)
$$
m_d\cong [m_sm_b]^{-1}|M(q_\downarrow)|,\eqno(4.15)
$$
$$
m_s-m_0(s)\cong -[m_s(m_b-m_s)]^{-1}|M(q_\downarrow)|\eqno(4.16)
$$
and
$$
m_b-m_0(b)\cong [m_b(m_b-m_s)]^{-1}|M(q_\downarrow)|.\eqno(4.17)
$$
Likewise, (4.11) leads to
$$
\beta_\downarrow\cong m_s\cos^2\theta_\downarrow,\eqno(4.18)
$$
$$
\alpha_\downarrow\cong (m_b-m_s\cos^2\theta_\downarrow)\cos^2
\phi_\downarrow\eqno(4.19)
$$
and
$$
m_d\cong 2m_s(1-\frac{m_s}{m_b}\cos^2\theta_\downarrow)
\sin^2\theta_\downarrow\cos^2\theta_\downarrow\sin^2\phi_\downarrow\cos^2\phi_\downarrow
(1-\cos{\displaystyle \chi}_{\scriptscriptstyle
T}(\downarrow)).\eqno(4.20)
$$
In (4.20), we may further neglect $m_s/m_b$ as compared to $1$;
this yields
$$
m_d\cong 2m_s\sin^2\theta_\downarrow\cos^2\theta_\downarrow
\sin^2\phi_\downarrow\cos^2\phi_\downarrow(1-\cos{\displaystyle
\chi}_{\scriptscriptstyle T}(\downarrow)).\eqno(4.21)
$$
When ${\displaystyle \chi}_{\scriptscriptstyle T}(\downarrow)=0$,
we
have $m_d=0$.\\

\noindent {\bf 4.2 Eigenstates of
$M(q_\downarrow)=M_0(q_\downarrow)+M_1(q_\downarrow)$}

Let $d,~s,~b$ be the normalized eigenstates of $M(q_\downarrow)$,
with
$$
M(q_\downarrow)|d)=m_d|d)
$$
$$
M(q_\downarrow)|s)=m_s|s)\eqno(4.22)
$$
$$
M(q_\downarrow)|b)=m_b|b).
$$
Throughout the paper, the $3\times 1$ normalized state vectors
$|d),~|s),~|b)$ may be denoted simply by $d,~s,~b$ as well.
Likewise, the states $\epsilon_\downarrow,~p_\downarrow$ and
$P_\downarrow$ of (3.8)-(3.10) may also be denoted by
$|\epsilon_\downarrow),~|p_\downarrow)$ and $|P_\downarrow)$.
Introduce the perturbation matrix
$$
g_\downarrow\cong (U_\downarrow)_0^\dag
M_1(q_\downarrow)(U_\downarrow)_0\eqno(4.23)
$$
by using $\epsilon_\downarrow,~p_\downarrow$ and $P_\downarrow$ as
base vectors, with $(U_\downarrow)_0$ given by (3.20). To the
lowest order in $\sin {\displaystyle \chi}_{\scriptscriptstyle
T}(\downarrow)$,
$$
g\equiv
g_\downarrow=i\alpha_\downarrow\xi_\downarrow\sin{\displaystyle
\chi}_{\scriptscriptstyle T}(\downarrow) \left(
\begin{array}{ccc}
~~0 ~~& 0 & \sin\theta_\downarrow\\
0 & 0 & \cos\theta_\downarrow\\
-\sin\theta_\downarrow &  \cos\theta_\downarrow& 0\\
\end{array}
\right) +O({\displaystyle \chi}_{\scriptscriptstyle T}^2).
\eqno(4.24)
$$
The corresponding eigenstates $d,~s,~b$ to the first order in
$\sin{\displaystyle \chi}_{\scriptscriptstyle T}(\downarrow)$ are
given by
$$
|d)=|\epsilon_\downarrow)-\frac{1}{\lambda(P_\downarrow)}
g_{{\scriptscriptstyle P}\epsilon}|P_\downarrow)
$$
$$
|s)=|p_\downarrow)+\frac{1}{m_s-\lambda(P_\downarrow)}g_{{\scriptscriptstyle
P}p}|P_\downarrow)\eqno(4.25)
$$
and
$$
|b)=|P_\downarrow)+\frac{1}{m_b}g_{\epsilon {\scriptscriptstyle
P}}|\epsilon_\downarrow)+\frac{1}{m_b-\lambda(p_\downarrow)}g_{p{\scriptscriptstyle
P}}|p_\downarrow)
$$
where
$$
g_{\epsilon {\scriptscriptstyle P}}=g^*_{{\scriptscriptstyle
P}\epsilon}=i\alpha_\downarrow\xi_\downarrow\sin\theta_\downarrow\sin{\displaystyle
\chi}_{\scriptscriptstyle T} (\downarrow)
$$
$$
{\sf
and}~~~~~~~~~~~~~~~~~~~~~~~~~~~~~~~~~~~~~~~~~~~~~~~~~~~~~~\eqno(4.26)
$$
$$
g_{p{\scriptscriptstyle P}}=g^*_{{\scriptscriptstyle
P}p}=i\alpha_\downarrow\xi_\downarrow\cos\theta_\downarrow\sin{\displaystyle
\chi}_{\scriptscriptstyle T}(\downarrow)
$$
are the first order nonzero matrix elements of $g_\downarrow$ in
accordance with (4.24).\\

\noindent {\bf 4.3 CKM Matrix and Jarlskog Invariant}

Anticipating that the Jarlskog Invariant $J$ in this model is
dominated by the $\downarrow$ quark sector because of
$$
\frac{m_um_c}{m_t^2}<<\frac{m_dm_s}{m_b^2}
$$
in accordance with (1.11), our discussions can be much simplified
by setting the $T$-violating phase
$$
{\displaystyle \chi}_{\scriptscriptstyle T}(\uparrow)=0\eqno(4.27)
$$
as an approximation. In this case,
$$
|u)\cong |\epsilon_\uparrow),~|c)\cong |p_\uparrow)~~{\sf
and}~~|t)\cong |P_\uparrow)\eqno(4.28)
$$
and
$$
U_\uparrow\cong
(U_\uparrow)_0=(\epsilon_\uparrow,~p_\uparrow,~P_\uparrow).\eqno(4.29)
$$
The corresponding CKM matrix is given by
$$
U_{CKM}=U_\uparrow^\dag U_\downarrow,\eqno(4.30)
$$
with its matrix elements in this approximation given by
$$
\begin{array}{lll}
U_{ud}=&u^\dag d=&\epsilon^\dag_\uparrow\epsilon_\downarrow
+i(\alpha\xi\sin\theta\sin{\displaystyle
\chi}_{\scriptscriptstyle T})_\downarrow\frac{\epsilon_\uparrow^\dag P_\downarrow}{m_b},\\
U_{cd}=&c^\dag d=&p^\dag_\uparrow\epsilon_\downarrow
+i(\alpha\xi\sin\theta\sin{\displaystyle
\chi}_{\scriptscriptstyle T})_\downarrow\frac{p_\uparrow^\dag P_\downarrow}{m_b},\\
U_{td}=&t^\dag d=&P^\dag_\uparrow\epsilon_\downarrow
+i(\alpha\xi\sin\theta\sin{\displaystyle
\chi}_{\scriptscriptstyle T})_\downarrow\frac{P_\uparrow^\dag P_\downarrow}{m_b},\\
U_{us}=&u^\dag s=&\epsilon^\dag_\uparrow p_\downarrow
+i(\alpha\xi\cos\theta\sin{\displaystyle
\chi}_{\scriptscriptstyle T})_\downarrow\frac{\epsilon_\uparrow^\dag P_\downarrow}{m_b-m_s},\\
U_{cs}=&c^\dag s=&p^\dag_\uparrow p_\downarrow
+i(\alpha\xi\cos\theta\sin{\displaystyle
\chi}_{\scriptscriptstyle T})_\downarrow\frac{p_\uparrow^\dag P_\downarrow}{m_b-m_s},\\
U_{ts}=&t^\dag s=&P^\dag_\uparrow p_\downarrow
+i(\alpha\xi\cos\theta\sin{\displaystyle \chi}_{\scriptscriptstyle
T})_\downarrow\frac{P_\uparrow^\dag P_\downarrow}{m_b-m_s},
\end{array}\eqno(4.31)
$$
etc., in which
$\epsilon^\dag_\uparrow\epsilon_\downarrow,~p^\dag_\uparrow\epsilon_\downarrow$,
etc. are given by the approximate matrix elements in $(U_{CKM})_0$
of (3.22).

Define
$$
{\cal S}_1\equiv U_{ud}^*U_{us}
$$
$$
{\cal S}_2\equiv U_{cd}^*U_{cs}\eqno(4.32)
$$
and
$$
{\cal S}_3\equiv U_{td}^*U_{ts}.
$$
We have
$$
{\cal S}_1+{\cal S}_2+{\cal S}_3=0\eqno(4.33)
$$
and the Jarlskog Invariant
$$
J=Im{\cal S}_1^*{\cal S}_2=Im{\cal S}_2^*{\cal S}_3=Im{\cal
S}_3^*{\cal S}_1.\eqno(4.34)
$$
Assume $\theta_\uparrow$ and $\theta_\downarrow$ are all small and
$O(\lambda)$, with $\lambda$ given by (3.25). To the lowest order
in powers of $\lambda=0.227$ and of $m_s/m_b$, we find
$$
J\cong \frac{m_s}{m_b}\sin(\theta_\uparrow-\theta_\downarrow)
\sin\theta_\downarrow\cos\theta_\downarrow\sin\phi_\downarrow\cos\phi_\downarrow
\sin\phi\sin{\displaystyle \chi}_{\scriptscriptstyle T}\eqno(4.35)
$$
where
$$
\phi=\phi_\uparrow-\phi_\downarrow.\eqno(4.36)
$$
Combining the square root of (4.21) with (4.35), we derive, to the
accuracy of the calculated order,
$$
J=\bigg(\frac{m_dm_s}{m_b^2}\bigg)^{1/2}\sin(\theta_\uparrow-\theta_\downarrow)
\sin\phi\cos[{\scriptstyle \frac{1}{2}}{\displaystyle
\chi}_{\scriptscriptstyle T}(\downarrow)].\eqno(4.37)
$$
By using (3.25)-(3.26), we derive (1.11) for $J$, which is
consistent with all available data. As the time reversal violating
phase ${\displaystyle \chi}_{\scriptscriptstyle
T}(\downarrow)\rightarrow 0$, both $J$ and $(m_dm_s/m_b^2)^{1/2}$
approach zero; their ratio remains fixed by the $T$-conserving
elements of the CKM matrix:
$$
\sin(\theta_\uparrow-\theta_\downarrow)\cdot
\sin\phi=A\lambda^2\cdot\lambda.\eqno(4.38)
$$
It is satisfying that this limiting value is consistent with
available experimental data, as shown by (1.11)-(1.13).

\newpage

\section*{\Large \sf  5. Lepton Sectors (neglecting $T$ violations)}
\setcounter{section}{5} \setcounter{equation}{0}

The application of hidden symmetry to the lepton sectors will be
examined in this and the following sections.\\

\noindent {\bf 5.1 General Discussion}

The lepton mass operators are given by (1.7)-(1.8). In the zeroth
approximation of $T$ invariance, these operators can be written as
$$
{\cal M}_0(l_\downarrow) = a_\downarrow | l_3(\downarrow) -
\kappa_\downarrow l_2(\downarrow)|^2 + b_\downarrow |
l_2(\downarrow) - \rho_\downarrow l_1(\downarrow)|^2 +c_\downarrow
|l_1(\downarrow) - \sigma_\downarrow l_3(\downarrow)|^2\eqno(5.1)
$$
and
$$
{\cal M}_0(l_\uparrow) = a_\uparrow | l_3(\uparrow) -
\kappa_\uparrow l_2(\uparrow)|^2 + b_\uparrow | l_2(\uparrow) -
\rho_\uparrow l_1(\uparrow)|^2 +c_\uparrow |l_1(\uparrow) -
\sigma_\uparrow l_3(\uparrow)|^2\eqno(5.2)
$$
where the twelve parameters
$a_\uparrow,~a_\downarrow,~\kappa_\uparrow,~\kappa_\downarrow,~b_\uparrow,~b_\downarrow,~\cdots$
are all real, with at least six of them
$a_\uparrow,~a_\downarrow,~b_\uparrow,~b_\downarrow,~c_\uparrow,~c_\downarrow$
positive. As in (2.2) and (3.6), we impose
$$
\kappa_\downarrow \rho_\downarrow \sigma_\downarrow =
\kappa_\uparrow \rho_\uparrow \sigma_\uparrow =1,\eqno(5.3)
$$
$$
\frac{b_\downarrow}{c_\downarrow}=\sigma_\downarrow^2~~{\sf and}~~
\frac{b_\uparrow}{c_\uparrow}=\sigma_\uparrow^2.\eqno(5.4)
$$
Hence, as in (4.2) these mass operators become
$$
{\cal M}_0(l_\downarrow) = a_\downarrow | l_3(\downarrow) -
\kappa_\downarrow l_2(\downarrow)|^2 + b_\downarrow |
l_2(\downarrow) - \rho_\downarrow l_1(\downarrow)|^2 +b_\downarrow
|l_3(\downarrow) - \kappa_\downarrow \rho_\downarrow
l_1(\downarrow)|^2\eqno(5.5)
$$
and
$$
{\cal M}_0(l_\uparrow) = a_\uparrow | l_3(\uparrow) -
\kappa_\uparrow l_2(\uparrow)|^2 + b_\uparrow | l_2(\uparrow) -
\varrho_\uparrow l_1(\uparrow)|^2 +b_\uparrow |l_3(\uparrow) -
\kappa_\uparrow \rho_\uparrow l_1(\uparrow)|^2.\eqno(5.6)
$$
Correspondingly, the mass matrices $M(l_\downarrow)$ and
$M(l_\uparrow)$ defined by (1.7)-(1.8) are, similar to (4.6),
$$
[M_0(l)]_{\downarrow~{\sf or}~\uparrow} = \left(
\begin{array}{ccc}
b\rho^2(1+\kappa^2) & -b\rho & -b\kappa\rho\\
-b\rho &b+a\kappa^2 & -a\kappa\\
-b\kappa\rho & -a\kappa & a+b
\end{array}
\right)_{\downarrow~{\sf or}~\uparrow}~. \eqno(5.7)
$$
Since their determinants satisfy
$$
|M_0(l_\downarrow)|=|M_0(l_\uparrow)|=0,\eqno(5.8)
$$
each matrix has an eigenvector of zero eigenvalue:
$$
M_0(l_\downarrow)\delta_\downarrow=0\eqno(5.9)
$$
and
$$
M_0(l_\uparrow)\delta_\uparrow=0.\eqno(5.10)
$$
As in (3.1)-(3.2), let $(V_\downarrow)_0$ and $(V_\uparrow)_0$ be
the real unitary matrices that diagonalize $M_0(l_\downarrow)$ and
$M_0(l_\uparrow)$:
$$
(V_\downarrow)_0^\dag M_0(l_\downarrow)(V_\downarrow)_0 =\left(
\begin{array}{ccc}
m_0(e)&0&0\\
0&m_0(\mu)&0\\
0&0&m_0(\tau)
\end{array}
\right )\eqno(5.11)
$$
and
$$
(V_\uparrow)_0^\dag M_0(l_\uparrow)(V_\uparrow)_0 =\left(
\begin{array}{ccc}
m_0(\nu_1)&0&0\\
0&m_0(\nu_2)&0\\
0&0&m_0(\nu_3)
\end{array}
\right ).\eqno(5.12)
$$
The corresponding zeroth order neutrino mapping matrix is
$$
(V_\nu)_0=(V_\downarrow)_0^\dag(V_\uparrow)_0.\eqno(5.13)
$$
Note that the roles of $\uparrow$ and $\downarrow$ in (5.13) are
switched in comparison between those in (3.3) because of our
accustomed definitions of $V_\nu$ and $U_{CKM}$.

We will further simplify the lepton mass operators by assuming
$$
\kappa_\downarrow~=-1,\rho_\downarrow=y,~\sigma_\downarrow=-\frac{1}{y}
$$
$$
~~~~~~~~~~~~~~~~~~~~~~~~~~~~~~~~~~~~~~~~~~~~\eqno(5.14)
$$
$$
\kappa_\uparrow~=x,\rho_\uparrow=-\frac{1}{\sqrt{2}}~{\sf
and}~~\sigma_\uparrow=-\frac{\sqrt{2}}{x}
$$
with $x,~y$ two small real parameters; i.e.,
$$
|x|<<1~~{\sf and}~~|y|<<1.\eqno(5.15)
$$
Thus, (5.1) and (5.2) become
$$
{\cal M}_0(l_\downarrow) = a_\downarrow | l_3(\downarrow) +
l_2(\downarrow)|^2 + b_\downarrow | l_2(\downarrow) - y
l_1(\downarrow)|^2 +b_\downarrow |l_3(\downarrow) +y
l_1(\downarrow)|^2\eqno(5.16)
$$
and
$$
{\cal M}_0(l_\uparrow) = a_\uparrow | l_3(\uparrow) - x
l_2(\uparrow)|^2 + b_\uparrow | l_2(\uparrow) +
 {\scriptstyle \sqrt{\frac{1}{2}}}~l_1(\uparrow)|^2 +b_\uparrow |l_3(\uparrow)
+{\scriptstyle \sqrt{\frac{1}{2}}}~x l_1(\uparrow)|^2.\eqno(5.17)
$$
Correspondingly (5.7) can be written as
$$
M_0(l_{\downarrow}) = \left(
\begin{array}{ccc}
2b_\downarrow y^2 & -b_\downarrow y & b_\downarrow y\\
-b_\downarrow y &a_\downarrow +b_\downarrow& a_\downarrow\\
b_\downarrow y & a_\downarrow & a_\downarrow +b_\downarrow
\end{array}
\right) \eqno(5.18)
$$
and
$$
M_0(l_{\uparrow}) = \left(
\begin{array}{ccc}
\frac{1}{2}~b_\uparrow (1+x^2) & \sqrt{\frac{1}{2}}~b_\uparrow & \sqrt{\frac{1}{2}}~b_\uparrow x\\
\sqrt{\frac{1}{2}}~b_\uparrow &b_\uparrow +a_\uparrow x^2 & -a_\uparrow x\\
\sqrt{\frac{1}{2}}~b_\uparrow x & -a_\uparrow x & a_\uparrow
+b_\uparrow
\end{array}
\right)~. \eqno(5.19)
$$
Since the matrices (5.18) and (5.19) are special cases of the
matrix (2.3) with the constraints (2.5) and (3.6), their
eigenvectors and eigenvalues can be readily obtained. Arrange the
three eigenvalues $\lambda_e,~\lambda_m$ and $\lambda_t$ of
$M_0(l_\downarrow)$ in the same order as those in (3.11)-(3.13)
and (5.11); we have
$$
m_0(e)=\lambda_e=0,\eqno(5.20)
$$
$$
m_0(\mu)=\lambda_m=b_\downarrow(1+2y^2)\eqno(5.21)
$$
and
$$
m_0(\tau)=\lambda_t=2a_\downarrow+b_\downarrow.\eqno(5.22)
$$
Likewise, the eigenvalues of $M_0(l_\uparrow)$, in the same
ascending order as (3.17)-(3.19) and (5.12), are
$$
m_0(\nu_1)=\lambda_n=0,\eqno(5.23)
$$
$$
m_0(\nu_2)=\lambda_l={\scriptstyle
\frac{1}{2}}~b_\uparrow(3+x^2)\eqno(5.24)
$$
and
$$
m_0(\nu_3)=\lambda_L=a_\uparrow(1+x^2)+b_\uparrow.\eqno(5.25)
$$
Similarly, by using (3.8)-(3.10) and (3.14)-(3.16) the
corresponding eigenvectors of $M_0(l_\downarrow)$ and
$M_0(l_\uparrow)$ can be readily written down.\\

{\bf 5.2 A limiting Case}

In the limit
$$
x\rightarrow 0~~{\sf and}~~y\rightarrow 0,\eqno(5.26)
$$
the mass matrices (5.18) and (5.19) become
$$
[M_0(l)]_{\downarrow} \rightarrow \left(
\begin{array}{ccc}
0 & 0 & 0\\
0 &a_\downarrow +b_\downarrow& a_\downarrow\\
0 & a_\downarrow & a_\downarrow +b_\downarrow
\end{array}
\right) \eqno(5.27)
$$
and
$$
[M_0(l)]_{\uparrow}\rightarrow \left(
\begin{array}{ccc}
\frac{1}{2}~b_\uparrow  & \sqrt{\frac{1}{2}}~b_\uparrow & 0\\
\sqrt{\frac{1}{2}}~b_\uparrow &b_\uparrow & 0\\
0 &0 & a_\uparrow +b_\uparrow
\end{array}
\right)~. \eqno(5.28)
$$
In accordance with (5.11)-(5.12), the matrices $(V_\downarrow)_0$
and $(V_\uparrow)_0$ becomes
$$
(V_\downarrow)_0 \rightarrow \left(
\begin{array}{ccc}
1 & 0 & 0\\
0 &\sqrt{\frac{1}{2}}&\sqrt{\frac{1}{2}}\\
0 & -\sqrt{\frac{1}{2}}& \sqrt{\frac{1}{2}}
\end{array}
\right) \eqno(5.29)
$$
and
$$
(V_\uparrow)_0 \rightarrow \left(
\begin{array}{ccc}
-\sqrt{\frac{2}{3}} & \sqrt{\frac{1}{3}} & 0\\
\sqrt{\frac{1}{3}} &\sqrt{\frac{2}{3}}&0\\
0 & 0& 1
\end{array}
\right), \eqno(5.30)
$$
with the corresponding neutrino-mapping matrix $(V_\nu)_0$ given
by the Harrison-Scott form
$$
(V_\nu)_0\equiv(V_\downarrow)_0^\dag(V_\uparrow)_0 \rightarrow
\left(
\begin{array}{ccc}
-\sqrt{\frac{2}{3}} & \sqrt{\frac{1}{3}} & 0\\
\sqrt{\frac{1}{6}} &\sqrt{\frac{1}{3}}&-\sqrt{\frac{1}{2}}\\
\sqrt{\frac{1}{6}} & \sqrt{\frac{1}{3}}& \sqrt{\frac{1}{2}}
\end{array}
\right). \eqno(5.31)
$$
Thus, the matrices (5.18) and (5.19) agree with all existing
experimental data provided $x$ and $y$ are small.

\newpage

\section*{\Large \sf  6. Lepton Sectors (with $T$ violation)}
\setcounter{section}{6} \setcounter{equation}{0}

With $T$ violations, we generalize (5.16)-(5.17) and write in
accordance with (1.7)-(1.8):
$$
{\cal
M}(l_\downarrow)=\bigg(\bar{l}_1(\downarrow),~\bar{l}_2(\downarrow),~\bar{l}_3(\downarrow)\bigg)
M(l_\downarrow) \left(
\begin{array}{r}
l_1(\downarrow)\\
l_2(\downarrow)\\
l_3(\downarrow)
\end{array}\right)
$$
$$
= a_\downarrow | l_3(\downarrow) +e^{i\phi_\downarrow}
l_2(\downarrow)|^2 + b_\downarrow | l_2(\downarrow) - y
l_1(\downarrow)|^2 +b_\downarrow |l_3(\downarrow) +y
l_1(\downarrow)|^2, \eqno(6.1)
$$
and
$$
{\cal
M}(l_\uparrow)=\bigg(\bar{l}_1(\uparrow),~\bar{l}_2(\uparrow),~\bar{l}_3(\uparrow)\bigg)
M(l_\uparrow) \left(
\begin{array}{r}
l_1(\uparrow)\\
l_2(\uparrow)\\
l_3(\uparrow)
\end{array}\right)
$$
$$
= a_\uparrow | l_3(\uparrow) - x l_2(\uparrow)|^2 + b_\uparrow |
l_2(\uparrow) +{\scriptstyle
\sqrt{\frac{1}{2}}}~e^{i\phi_\uparrow} l_1(\uparrow)|^2
+b_\uparrow |l_3(\uparrow) +{\scriptstyle \sqrt{\frac{1}{2}}}~x
l_1(\uparrow)|^2.\eqno(6.2)
$$
The matrices $M(l_\downarrow)$ and $M(l_\uparrow)$ can also be
written as
$$
M(l_\downarrow)=M_0(l_\downarrow)+M_1(l_\downarrow)\eqno(6.3)
$$
and
$$
M(l_\uparrow)=M_0(l_\uparrow)+M_1(l_\uparrow)\eqno(6.4)
$$
with $M_0(l_\downarrow)$ and $M_0(l_\uparrow)$ given by (5.18)
and(5.19),
$$
M_1(l_{\downarrow}) =a_\downarrow \left(
\begin{array}{ccc}
0 & 0 & 0\\
0 &0& e^{-i\phi_\downarrow}-1\\
0 & e^{i\phi_\downarrow}-1& 0
\end{array}
\right) \eqno(6.5)
$$
and
$$
M_1(l_{\uparrow}) =\frac{1}{\sqrt{2}}~b_\uparrow \left(
\begin{array}{ccc}
0 &  e^{-i\phi_\uparrow}-1 & 0\\
e^{i\phi_\uparrow}-1&0& 0\\
0 & 0& 0
\end{array}
\right) \eqno(6.6)
$$
in which $\phi_\downarrow$ and $\phi_\uparrow$ are the
$T$-violating phases.

The determinants of $M(l_\downarrow)$ and $M(l_\uparrow)$ are
$$
|M(l_\downarrow)|=2a_\downarrow
b_\downarrow^2y^2(1-\cos\phi_\downarrow)\eqno(6.7)
$$
and
$$
|M(l_\uparrow)|=a_\uparrow
b_\uparrow^2x^2(1-\cos\phi_\uparrow).\eqno(6.8)
$$
The three eigenvalues of $M(l_\downarrow)$ are the physical masses
of charged leptons $e,~\mu$ and $\tau$:
$$
m=m_e,~m_\mu,~m_\tau,\eqno(6.9)
$$
determined by the cubic equation
$$
m(m-\lambda_m)(m-\lambda_t)=|M(l_\downarrow)|\eqno(6.10)
$$
with $\lambda_m,~\lambda_t$ given by (5.21) and (5.22). Likewise,
the eigenvalues of $M(l_\uparrow)$ are the three neutrino masses
$$
m=m_1,~m_2,~m_3,\eqno(6.11)
$$
determined by
$$
m(m-\lambda_l)(m-\lambda_L)=|M(l_\uparrow)|\eqno(6.12)
$$
with $\lambda_l,~\lambda_L$ given by (5.24)-(5.25). As in (4.9),
Eqs. (6.10) and (6.12), together with (5.20)-(5.25) give a simple
way to derive the physical masses of leptons with the inclusion of
$T$ violation effects.

On the other hand, for evaluation of eigenvectors it is more
convenient to transform
$$
l_1(\downarrow)=l_1'(\downarrow),~~l_2(\downarrow)=e^{-i\phi_\downarrow/2}l_2'(\downarrow)~~{\sf
and}~~l_3(\downarrow)=e^{i\phi_\downarrow/2}l_3'(\downarrow).\eqno(6.13)
$$
Thus,
$$
{\cal M}(l_\downarrow)=
\bigg(\bar{l}'_1(\downarrow),~\bar{l}'_2(\downarrow),~\bar{l}'_3(\downarrow)\bigg)
\bigg(H_0(l_\downarrow)+H_1(l_\downarrow)\bigg) \left(
\begin{array}{r}
l'_1(\downarrow)\\
l'_2(\downarrow)\\
l'_3(\downarrow)
\end{array}\right)\eqno(6.14)
$$
with
$$
H_0(l_{\downarrow}) = \left(
\begin{array}{ccc}
2b_\downarrow y^2 & 0 & 0\\
0 &a_\downarrow +b_\downarrow&a_\downarrow\\
0 &a_\downarrow&a_\downarrow +b_\downarrow
\end{array}\right)\eqno(6.15)
$$
and
$$
H_1(l_{\downarrow}) =b_\downarrow y \left(
\begin{array}{ccc}
0& -z_\downarrow^* & z_\downarrow\\
-z_\downarrow &0&0\\
z_\downarrow^* &0&0
\end{array}
\right), \eqno(6.16)
$$
with
$$
z_\downarrow=e^{i\frac{1}{2}\phi_\downarrow}.\eqno(6.17)
$$
Likewise, introduce $l_1'(\uparrow)$ and $l_2'(\uparrow)$ through
$$
l_1(\uparrow)=e^{-i\phi_\uparrow}l_1'(\uparrow),~~
l_2(\uparrow)=e^{i\phi_\uparrow}l_2'(\uparrow)~~{\sf
and}~~l_3(\uparrow)=l_3'(\uparrow).\eqno(6.18)
$$
Correspondingly,
$$
{\cal M}(l_\uparrow)=
\bigg(\bar{l}'_1(\uparrow),~\bar{l}'_2(\uparrow),~\bar{l}'_3(\uparrow)\bigg)
\bigg(H_0(l_\uparrow)+H_1(l_\uparrow)\bigg) \left(
\begin{array}{r}
l'_1(\uparrow)\\
l'_2(\uparrow)\\
l'_3(\uparrow)
\end{array}\right)\eqno(6.19)
$$
with
$$
H_0(l_{\uparrow}) = \left(
\begin{array}{ccc}
\frac{b_\uparrow}{2}(1+ x^2) & \sqrt{\frac{1}{2}}~b_\uparrow & 0\\
\sqrt{\frac{1}{2}}~b_\uparrow &b_\uparrow +a_\uparrow x^2&0\\
0 &0&a_\uparrow +b_\uparrow
\end{array}\right)\eqno(6.20)
$$
and
$$
H_1(l_{\uparrow}) =x\left(
\begin{array}{ccc}
0& 0& \sqrt{\frac{1}{2}}~ b_\uparrow z_\uparrow\\
0 &0&-a_\uparrow z_\uparrow^*\\
 \sqrt{\frac{1}{2}}~ b_\uparrow z_\uparrow^* &-a_\uparrow z_\uparrow&0
\end{array}
\right), \eqno(6.21)
$$
with
$$
z_\uparrow=e^{i\frac{1}{2}\phi_\uparrow}.\eqno(6.22)
$$
The smallness of $x$ and $y$ makes it possible to use
$H_1(\downarrow)$ and $H_1(\uparrow)$ as perturbations.

Let $V_\downarrow$ and $V_\uparrow$ be the unitary matrices that
diagonalize $M(l_\downarrow)$ and $M(l_\uparrow)$ of (6.1) and
(6.2).The corresponding neutrino mapping matrix is
$$
V_\nu=V_\downarrow^\dag V_\uparrow.\eqno(6.23)
$$
Define
$$
T_1=(V_\nu)_{21}^*(V_\nu)_{31},
$$
$$
T_2=(V_\nu)_{22}^*(V_\nu)_{32}\eqno(6.24)
$$
and
$$
T_3=(V_\nu)_{23}^*(V_\nu)_{33}.
$$
Their sum satisfies
$$
T_1+T_2+T_3=0.\eqno(6.25)
$$
The corresponding Jarlskog invariant for leptons is
$$
{\cal J}_\nu=ImT_1^*T_2=ImT_2^*T_3=ImT_3^*T_1.\eqno(6.26)
$$

To first order in $x$ and $y$, we find
$$
{\cal J}_\nu=-\frac{y}{3\sqrt{2}}\bigg[\cos
\frac{\phi_\downarrow}{2}\sin(\frac{\phi_\downarrow}{2}+\phi_\uparrow)+
\frac{b_\downarrow}{2a_\downarrow+b_\downarrow}\sin
\phi_\downarrow\cos(\frac{\phi_\downarrow}{2}+\phi_\uparrow)\bigg]
$$
$$
+\frac{x}{3}~\frac{a_\uparrow b_\uparrow}{(a_\uparrow
+b_\uparrow)(2a_\uparrow-b_\uparrow)}
\bigg[-\sin\phi_\downarrow+\sin(\phi_\downarrow+\phi_\uparrow)\bigg].\eqno(6.27)
$$
As in(4.37), there is an interesting relation between ${\cal
J}_\nu$ and lepton masses, which will be discussed in a separate
paper.

\section*{\Large \sf References}


\noindent [1] R. Friedberg and T. D. Lee, HEP \& NP 30(2006)591.\\

\noindent [2] P. F. Harrison and W. G. Scott, Phys. Lett. B535, 163(2002).\\

\noindent [3] L. Wolfenstein, Phys. Rev. D18, 958(1978);

P. F. Harrison, D. H. Perkins and W. G. Scott, Phys. Lett. B530,
167(2002);

Z. Z. Xing, Phys. Lett. B533, 85(2002);

X. G. He and A. Zee, Phys. Lett. B560, 87(2003).\\

\noindent [4] S. Eidelman et al. (Particle Data Group), Phys.
Lett. B592, 1(2004).\\

\newpage

{\normalsize \sf

\section*{\Large \sf Appendix}

In (2.3), the $3\times 3$ symmetric matrix is written in the form
$$
M = \left(
\begin{array}{ccc}
\gamma+\beta\eta^2 & -\beta\eta & -\gamma\zeta\\
-\beta\eta &\beta + \alpha\xi^2 & -\alpha\xi\\
-\gamma\zeta& -\alpha\xi & \alpha+\gamma\zeta^2
\end{array}
\right) \eqno(A.1)
$$
with six parameters $\alpha,~\beta,~\gamma,~\xi,~\eta$ and
$\zeta$. On the other hand, any such symmetric $M$ can also be
expressed by
$$
M = \left(
\begin{array}{ccc}
c & -e & -f\\
-e &b & -d\\
-f& -d & a
\end{array}
\right)\eqno(A.2)
$$
with
$$
a=\alpha+\gamma\zeta^2,\eqno(A.3)
$$
$$
b=\beta+\alpha\xi^2,\eqno(A.4)
$$
$$
c=\gamma+\beta\eta^2,\eqno(A.5)
$$
$$
d=\alpha\xi,\eqno(A.6)
$$
$$
e=\beta\eta\eqno(A.7)
$$
and
$$
f=\gamma\zeta.\eqno(A.8)
$$
In this Appendix, we give the answer to the inverse problem: Given
$a,~b,~\cdots,~f$ what are the corresponding
$\alpha,~\beta,~\cdots,~\zeta$?

Introduce
$$
A=bc-e^2\eqno(A.9)
$$
$$
B=ae^2+bf^2-cd^2-abc\eqno(A.10)
$$
and
$$
C=(ac-f^2)d^2.\eqno(A.11)
$$
One can readily verify that $\alpha$ satisfies
$$
A\alpha^2+B\alpha+C=0.\eqno(A.12)
$$
Knowing $\alpha$, (A.6) gives $\xi$. From $\alpha\xi^2$, $\beta$
can be determined. Thus, $\eta$ is known from (A.7). Likewise,
$\gamma$ can be deduced from (A.5) and $\zeta$ from (A.3).

For $\alpha,~\beta,~\gamma,~\xi,~\eta,~\zeta$ real, so are
$a,~b,~c,~d,~e,~f$. However, the converse is not always true, as
can be readily studied by examining the solutions of the quadratic
equation (A.12).

\end{document}